\documentclass[aps,pra,floatfix,showpacs,twocolumn,tightenlines,superscript,address]{revtex4}

\usepackage{amsmath}
\usepackage{graphicx}
\usepackage{color}
\usepackage{bm}
\usepackage{epsfig}
\usepackage{amsfonts}
\usepackage{amsthm}

\begin{document}

\newlength{\defbaselineskip}
\setlength{\defbaselineskip}{\baselineskip}
\newcommand{\interlinea}[1]{\setlength{\baselineskip}{#1 \defbaselineskip}}

\newcommand\ket[1]{\left|#1\right\rangle}
\newcommand\bra[1]{\left\langle#1\right|}
\newcommand\lavg{\left\langle}
\newcommand\ravg{\right\rangle}
\newcommand\be{\begin{equation}}
\newcommand\ee{\end{equation}}
\newcommand\la{\leftarrow}
\newcommand\ra{\rightarrow}
\newcommand\mf{\mathbf}
\newcommand\trm{\textrm}
\newcommand\id{{\rm 1} 
\hspace{-1.1mm} {\rm I}
\hspace{0.5mm}}
\newcommand\bea{\begin{eqnarray}}
\newcommand\eea{\end{eqnarray}}
\newcommand\stocavg[1]{{\lavg #1 \ravg}_{\trm{stoc}}}
\newcommand\fk[1]{\mathfrak{#1}}

\title{Yukawa bosons in two-dimensional harmonic confinement}

\author{K. K. Rajagopal}
\email[]{rajagoku@physics.uq.edu.au}
\affiliation{Department of Physics, The University of Queensland, 4072 Brisbane, Australia}
\date{15 February 2007}
\begin{abstract}  
The ground state property of Yukawa Bose fluid confined in a radial harmonic trap is studied. The calculation were carried out using the Density Functional Theory formalism within the Kohn-Sham scheme. The excess correlation energy for this inhomegeneous fluid is approximated $via$ the Local Density Approximation. A comparison is also made with the Gross-Piteavskii model. We found that the system of bosons interacting in term of Yukawa potential in a harmonic trap is energetically favourable compared to the ones interacting {\it via} contact delta potential.   
\end{abstract}

\maketitle

\section{Yukawa Bose Fluid, an introduction}

A system of $N$ Bose particles interacting via the Yukawa potential is called Yukawa Bose fluid (for short YBF). For three dimensional-space $D=3$ the Yukawa potential takes the form 
$V_{3}(\tilde{r})$=$(\epsilon\sigma/\tilde{r})\exp(-\tilde{r}/\sigma)$, where $\epsilon$ is an energy scale and $\sigma$ is a length scale having the meaning of a screening length. Whereas for the 
two-dimensional case however, it is described by $V_{2}(\tilde{r})=\epsilon K_{0}(\tilde{r}/\sigma)$ where $K_{0}(\tilde{x})$ is the modified Bessel function that decays as $\exp(-\tilde{x})/\sqrt{\tilde{x}}$ at large distance while it behaves in a repulsive manner as $-\ln(\tilde{x})$ at short distances. In other word the Yukawa potential presents a combination of short range with soft-core type potential.

Early numerical calculation (quantal Monte Carlo simulations) in studying the ground state property of the three-dimensional YKF can be traced back to the work of  Ceperley and coworkers \cite{ceperley1,ceperley2}. There were a spurt of interest in numerical calculation of 2D ground state property of the 2D-YBF in the 90's following the idea of Nelson and Seung \cite{nelson} who have shown that the statistical mechanics of the flux-line lattice (FLL) of high-$T_{c}$ superconductors can be studied through an appropriate mapping onto the 2D-YBF. Magro and Ceperley \cite{magro} have performed a Diffusion Monte Carlo (DMC) and Variational Monte Carlo (VMC) numerical simulation to calibrate the ground state vortex properties and phases (liquid or solid) that correspond to the density and kinetic energy of the system. Similar work related to the first order phase transition of Abrikosov lattice to liquid of vortices have been reported by Nordborg and Blatter \cite{blatter} using the path integral Monte Carlo method. In the late 90's other methods such as the STLS model \cite{tanatar,streppa} appeared in studying the ground state properties of the 2D-YBF. It has to be noted that all these works were spiralling around the system of homogeneous fluid. 

\begin{figure}
\centering{
\epsfig{file=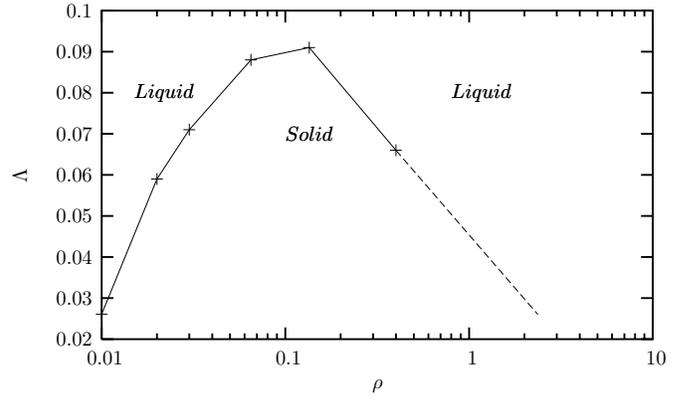,width=1.0\linewidth}}
\caption{The phase diagram of Yukawa bosons reproduced from Magro and 
Ceperley \cite{magro}. The pluses are transition points computed with DMC. 
The dashed line at high densities is the scaling law: 
$\Lambda\sim 0.04/\sqrt{\rho}$. }  
\label{fig_phase}
\end{figure}

Our work will focus on the harmonically trapped system extending the work based on the quantal Monte Carlo calculations of Magro and Ceperley \cite{magro}. Of particular interest to us is the phase diagram obtained by them for the parameters ($\rho$, $\Lambda$) at transition points displayed in Fig. (\ref{fig_phase}).  Here $\Lambda$ indicates the De Boer dimensionless parameter defined by $\Lambda^{2}=\hbar^2/2m\sigma^2\epsilon$ while $\rho$ and $m$ denote the reduced density and atomic mass respectively. 
 
Magro and Ceperley \cite{magro} observed that for a particular region $\Lambda > 0.09$ the system is dominated by kinetic energy and does not crystallize. Below this threshold however a peculiar behaviour of reentrant liquid has been observed. Which means that system is in the liquid phase at very low density as well at high density. The crystal melts on compression and expansion. 
It is known that at low density the system is in liquid phase due to the fact that the screened potential cannot bind to solid phase particles that are on average too far apart. For the high density the crystal is known to melt similar to the mechanism of thermal melting of a Wigner crystal. However in this work we are dealing with the liquid phase of Yukawa bosons only. 
Our calculation rely on the data of the liquid phase near the transition point from the phase diagram  Fig. (\ref{fig_phase}) as an input to the Local Density Approximation (LDA). The use of Density Functional theory (DFT)-LDA in this work is only a matter of computational convenience primarily due to the availability of the exact Monte-Carlo \cite{magro} data of the homogeneous system. Infact to our knowledge this is the first attempt to study the YBF in a trapped system.    

The motivation behind this work is that we believe the theory of 2D-YBF can be used in studying the ground state property of the strongly correlated liquid formed at the melted phase of many vortex system in the dilute trapped Bose gas \cite{cooper,sinova,fischer}. In the latter system Bose-Einstein condensate (BEC) is completely depleted by quantum fluctuations, and quantum liquids appear with excitations that can carry fractional statistics. In this situation vortices can be treated as fundamental objects forming themselves as strongly correlated quantum fluid. This idea has been used in explaining certain features of Hall conductance and magnetization experiments in high-$T_{c}$ superconductors \cite{blatter,horovitz,rozhkov}.
On the other hand, the repulsive nature of vortices interacting in the short ranges is well described by the soft-core logarithmic potential. At the edge of the trap the potential is smoothed out by $\exp(-\tilde{x})/\sqrt{\tilde{x}}$ that decays to zero making it computationaly tractable and rule out any instability that may occur at the trap boundary. These characteristic making it plausible for the vortex liquid to be mapped as Yukawa bosons in a similar fashion to one adapted by Nelson and Seung \cite{nelson} for the YBF-FLL in the high $T_{c}$ superconductors. 

The paper is outlined as follows. In section \ref{DFT} we introduce the Density functional Theory formalism in showing that by choosing a particular type of auxiliary external trapping potential the non-interacting system of bosons can be mapped to a real interacting system. We present the numerical calculation the Kohn-Sham equation and Gross-Pitaevskii (GP) in section \ref{Algorithm}. Then the numerical results will be discussed and concluded.\par

\section{Application of DFT theory in the Yukawa bosons system}
\label{DFT}

The Density functional theory (DFT) which is originally based on the notion that for a many-electron system there is a one-to-one mapping between the external potential and the electron density: $v_{ext}({\bf r})$ $\leftrightarrow$ $\rho({\bf r})$. In other words, the density is uniquely determined given a potential, and $vice\,\,versa$. All properties are therefore a functional of the density, because the density determines the potential, which determines the Hamiltonian, which determines the energy and 
the wave function. Following this train of reasoning, the inhomogeneous dilute system of $N$ 
interacting bosons can be described within the second quantization language as

\begin{eqnarray}
\hat{H} &=& \hat{H_{0}}+\int\,d{\bf r} \psi^\dagger({\bf r})V_{ext}({\bf r})
\psi({\bf r})\nonumber \\
&+&\frac{1}{2}\int\,d{\bf r}\int\,d{\bf r'}
\psi^\dagger({\bf r})\psi^\dagger({\bf r'})V(|{\bf r}-{\bf r'}|)
\psi({\bf r'})\psi({\bf r})\nonumber \\
&=& \hat{H_{0}}+\hat{V}_{ext}+\hat{V}_{int}
\label{TotHamil}
\end{eqnarray}    

where $\hat{H_{0}}=\int\,d{\bf r} \psi^\dagger({\bf r})\left[-\frac{\hbar^2}{2m}\nabla^2 -\mu \right ]\psi({\bf r})$ and $V(|{\bf r}-{\bf r'}|)$ is the inter-atomic interaction potential. Here $V_{ext}({\bf r})$ is the external trapping potential while $m$ and $\mu$ are the atomic mass and the chemical potential respectively. The annihilation and creation field operators are denoted by $\psi^\dagger({\bf r})$ and $\psi({\bf r'})$ respectively and obey Bose-Einstein commutation relations:
\begin{eqnarray}
&[\psi({\bf r}),\psi^{\dagger}({\bf r'})]=\delta({\bf r}-{\bf r'}) \nonumber \\
& [\psi({\bf r}),\psi({\bf r'})]= [\psi^{\dagger}({\bf r}),\psi^{\dagger}({\bf r'})]=0 \,.
\label{commuta}
\end{eqnarray}
Let us denote the ground state of the system as $\ket{\Psi_o}$ so the ground state energy is defined as $E_{o}=\bra{\Psi_o}\hat{H}\ket{\Psi_o}$ and the ground state density by 
$n_{o}({\bf r})=\bra{\Psi_o}\psi^{\dagger}\psi\ket{\Psi_o}$. The Hohenberg-Kohn (HK) theorem \cite{HK64} guarantees that there exists a unique functional of the density, 

\begin{equation}
F[n]= \hat{H_{0}}[n]+\hat{V}_{int}[n] \,,
\label{Fenergy}
\end{equation}

irrespective of the external potential. The theorem was originally proved for 
fermions but its generalization also covers bosons. Following HK, we can write 
the total energy functional of the system as following,

\begin{equation}
E[n]= F[n]+ \int d{\bf r}V_{ext}({\bf r})n({\bf r}) \,.
\label{Tenergy}
\end{equation}

Determination of the ground state energy $E_{o}$ follows by imposing the stationary 
conditions

\begin{equation}
\frac{\delta E[n]}{\delta n({\bf r})}=0
\label{vari-Primo}
\end{equation} 

where we will obtain the ground state density $n_{o}(r)$ that is uniquely determined by the choice of our external potential $V_{ext}$.  In general, the Hohenberg-Kohn theorem does not provide us with a computational scheme to determine the ground state energy. This is provided by the Kohn-Sham (KS) procedure \cite{KS65}. The idea is to use an auxiliary system (non-interacting reference system) and look for and external potential $V_{ext}^{s}$ such that the noninteracting has the same ground state density as the real, interacting system. We write the Hamiltonian of the auxiliary system in the following form:

\begin{eqnarray}
\hat{H}^{s}&=& \int d{\bf r}\, \psi^\dagger({\bf r})\left[-\frac{\hbar^2}{2m}
\nabla^2 -\mu \right ]\psi({\bf r}) \nonumber \\
&+&\int d{\bf r}\, \psi^\dagger({\bf r})V_{ext}^{s}({\bf r})\psi({\bf r})\,.
\label{AuxHamil}
\end{eqnarray}

Thus for the ground state $\ket{\Psi_{o}}$ we can define the unique total ground state energy functional of the auxiliary can be written as

\begin{equation}
E^{s}[n_{s}]= F^{s}[n_{s}]+ \int d{\bf r}\, n_{s}({\bf r})V_{ext}^{s}({\bf r})\,.
\label{AuxEnergy}
\end{equation}

By the KS scheme we note that  $E^{s}[n_{s}]$ can be approximated to $E[n]$ or in other word the density of the auxiliary  system $n_{s}({\bf r})$ is equivalent to the real system 
$n({\bf r})$ by choosing a proper choice of auxiliary external potential $V_{ext}^{s}$.\par

Using the above argument and comparing the energy term $F^{s}[n_{s}]$ of 
Eq. (\ref{AuxEnergy}) with Eqs. (\ref{Fenergy}), we can deduce the following functional relation:

\begin{equation}
  F[n]= F^{s}[n]+ V_{H}[n]+ F_{xc}[n]\,,
\label{Frelation}
\end{equation}

where the second term $\hat{V}_{H}$ is called the Hartree-energy defined as  

\begin{equation}
\hat{V}_{H}=\frac{1}{2}\int d{\bf r}\,\int d{\bf r'}\,V(|{\bf r}-{\bf r'}|)
 n({\bf r})n({\bf r'}) \,.
\label{Hartree}
\end{equation}
 
The last term in Eq. (\ref{Frelation}) represents the exchange-correlation 
energy $E_{xc}[n({\bf r})]$ that includes all the contributions to the 
interaction energy beyond mean field Hatree term. Calculating the variational derivatives 
in Eq. (\ref{vari-Primo}) using Eq.(\ref{Frelation}), one finds

\begin{equation}
\frac{\delta F^{s}[n]}{\delta n({\bf r})}+V_{H}({\bf r})+
\frac{\delta F_{xc}[n]}{\delta n({\bf r})}+V_{ext}({\bf r})=0
\label{vari-ultimo}
\end{equation}

where the Hartree field read 
\begin{equation}
V_{H}({\bf r})=\int d{\bf r'}\, V(|{\bf r}-{\bf r'}|)n({\bf r'})\,.
\label{Hart2}
\end{equation}
Performing a similar variational calculation on Eq. (\ref{AuxEnergy}) we 
deduce that the density of the auxiliary system is identical to the 
actual system if 
\begin{equation}
V_{ext}^{s}({\bf r})=V_{ext}({\bf r})+ V_{H}({\bf r})+ V_{xc}({\bf r})\,.
\label{Auxtrap}
\end{equation}
 
where we have introduced in Eq(\ref {Auxtrap}) the exchange-correlation potential 
$V_{xc}=\delta F_{xc}[n({\bf r})]/\delta n({\bf r})$ which is unknown for most of the system of interest and thus one has to resort to approximations such as the Local Density Approximation (LDA). This will be dealt with in the following subsequent section.

\subsection{The local density approximation}

 Before we can actually implement the Kohn-Sham formalism, we have to devise some workable approximation for the exchange-correlation potential $V_{xc}({\bf r})$. The first such approximation to be suggested was the Local Density Approximation, or LDA. The idea behind the LDA is very simple; it just ignores the non-local aspects of the functional dependence of 
$V_{xc}({\bf r})$. The true form of $V_{xc}(\bf{r})$ will depend not only on the local density $n({\bf r})$ but also on $n$ at all other points ${\bf r}'$ and this functional dependence is in general not known. This difficulty is avoided with the assumption that $V_{xc}$ depends only on the local density $n({\bf r})$, and that $E_{xc}[n]$ can thus be written as

\begin{equation}
 E_{xc}[n({\bf r})]\approx \int\,d{\bf r} E_{xc}^{hom}[\rho]|_{\rho\rightarrow n({\bf r})}\,
\label{LDA}
\end{equation}

where $E_{xc}^{hom}[\rho]=\rho\epsilon_{xc}[\rho]$  and $\epsilon_{xc}[\rho]$ 
is the exchange correlation energy of a homogeneous system with uniform 
density $\rho$. The Functional derivatives of the above relation reads :
\begin{equation}
V_{xc}[n({\b r})]=\frac{\delta E_{xc}[n({\bf r})]}{\delta  n({\bf r})}=
\frac{\partial (\rho \epsilon_{xc}[\rho])}{\partial \rho}|_{\rho\rightarrow 
n({\bf r})}\,.
\label{partial}
\end{equation}

The available numerical data  of Magro and Ceperley \cite{magro} and 
Strepparola et. al \cite{streppa} permit one to obtain information on the 
homogeneous excess free energy $ f_{ex}[n]$ rather than the homogeneous 
exchange correlation energy $ f_{xc}[n]$ . Thus it is much convenient to work 
with excess free energy defined as \cite{moroni}

\begin{equation}
F_{ex}[n({\bf r})]= \hat{V}_{H}[n({\bf r})]+F_{xc}[n({\bf r})] \,.
\label{excess}
\end{equation}

In general, the excess correlation functional energy  $F_{ex}[n({\bf r)}]$ 
in a DFT calculation is not known exactly. One can resort to approximations 
such as the Local density approximation (LDA) which reads,

\begin{equation}
 F_{ex}[n({\bf r})]\approx \int\,d{\bf r} F_{ex}^{hom}[n]|_{
n\rightarrow n({\bf r})}\,
\label{LDA}
\end{equation}

where $F_{ex}^{hom}[n]=n f_{ex}^{hom}[n]$. The functional derivatives 
(excess-correlation potential) of the above relation can be 
written as: 

\begin{equation}
V_{ex}({\bf r},n({\bf r}))=\frac{\delta F_{ex}[n({\bf r})]}{\delta  n({\bf r})}
=\frac{\partial (n f_{ex}[n]) }{\partial n}|_{n\rightarrow 
n({\bf r})}\,.
\label{partial}
\end{equation}

\begin{figure}
\centering{
\epsfig{file=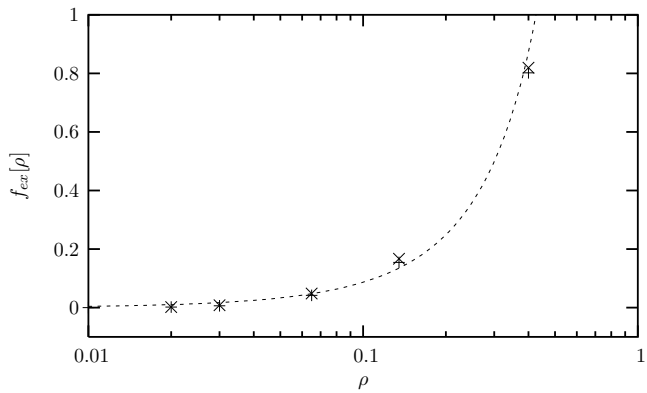,width=1.0\linewidth}}
\caption{Homogeneous excess-correlation energy $f_{ex}(\rho)$ versus the dimensionless homogenous gas density $\rho$ of the VMC data of Magro and Ceperley \cite{magro} (pluses) and STLS data of Strepparola et. al. \cite{streppa} (crosses) compared to the fit function 
Eq. (\ref{gfit}) (dashed line).}  
\label{fig_fit}
\end{figure}

Information of the homogeneous excess correlation energy $f_{ex}[n]$ can be 
obtained by subtracting the kinetic energy from the total ground state energy
of a homogeneous system with N boson. In Fig. (\ref{fig_fit}) a plot of the DMC data of 
Magro and Ceperley \cite{magro} and the STLS data of Strepparola et. al. \cite{streppa} are depicted along with the following fit:  

\begin{equation} 
f_{ex}[\rho]= -\frac{2\rho}{\log(\rho)}\,.
\label{gfit} 
\end{equation} 

\section{Numerical calculation for the Kohn-Sham equation.} 
\label{Algorithm}

We summarize here the algorithm in our numerical calculations which are commonly used in the literature to minimize energy functional $E[\Psi,\Psi^{*}]=<\Psi|H_{o}+V_{ext}^{s}|\Psi>$ where $\Psi=\sqrt{n(r)}\exp(i\phi)$ in which the phase $\phi$ fixes the velocity of the fluid through the relation $v=\nabla \phi/m$ . We minimize $ E[\Psi,\Psi^{*}]$ by assuming the normalization condition $\int dr \Psi^{*}\Psi=N$, to obtain the non-linear the time independent non-linear Kohn-Sham equation :

\begin{equation}
 \left[-\nabla^{2}+ x^{2}-\frac{2 n(x)}{\log[n(x)]}\right ]\Psi(x)=\mu\Psi(x) \, . 
\label{Kohn-Sham}
\end{equation}

In the above equation we have incorporated the auxiliary external potential based on Eq.(\ref{Auxtrap}) and Eqs. (\ref{excess})- (\ref{gfit}) using an isotropic planar trapping external potential $V_{ext}=1/2m\omega^{2}r^{2}$ with $\omega$ as radial 
frequency. We have scaled all length and energy by harmonic oscillator unit $a_{ho}=\sqrt{\hbar/m\omega}$ and 
$\hbar\omega/2$ respectively. We obtained the ground state solution of the system by numerical iteration method by using a 
two-step Crank-Nicholson discretization technique. 
 
\begin{figure}
\centering{
\epsfig{file=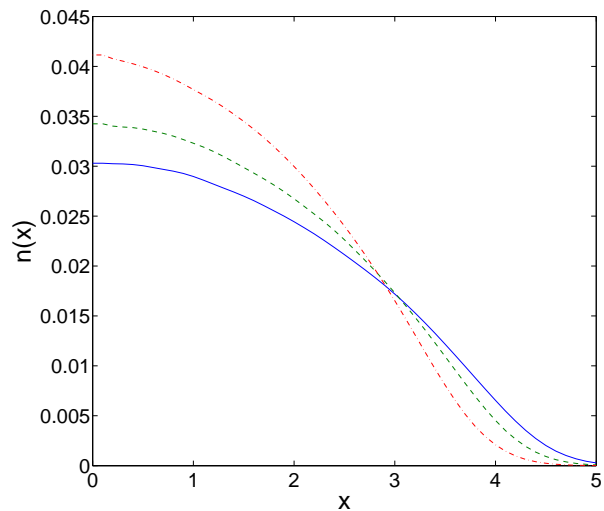,width=1.0\linewidth}}
\caption{Top panel : Density profiles of the YBF $n(x)$ (in arbitrary units) as a function of $x$ (in $\sqrt{\hbar/m\omega}$ units ) for various number of atoms $N=$ 200 (solid line), 150 (dashed line) and 100 (dashed dot line).} 
\label{figure3}
\end{figure}

The density profiles as a function of $N$ is shown in Fig. (\ref{figure3}). The profile shows a maximum at the center of trap and it decreases monotonically with $x$. To make a quantitative measurement, we fix the solution for Eq. (\ref{Kohn-Sham}) for $N=100$ ($n_{1}$) as a standard and compare the differences in density for the other two values at the trap center 
($\Delta n = n_{2}(0)-n_{1}(0)$ and $\Delta \tilde{n} = n_{3}(0)-n_{1}(0)$ ). Here $n_{2}(0)$ and $n_{2}(0)$ corresponds to the solution of Eq. (\ref{Kohn-Sham}) for $N=150$ and $N=200$ respectively at the center of trap. We found the differences $\Delta n$ and $\Delta \tilde{n}$ decreases with $ \Delta n/n_{1}$ and  $\Delta \tilde{n}/n_{1}$ by as large as 17$\%$ to 26$\%$. The explaination of these results is quite straight forward. The areal integral of the density $n(r)$ yields the total number of atoms in the system. Hence larger number of atoms produced a much profound profile. The central density of the cloud decreases rapidly with increasing $N$ but the density distribution is flattened due to stronger repulsion between particles. 

\begin{figure}
\centering{
\epsfig{file=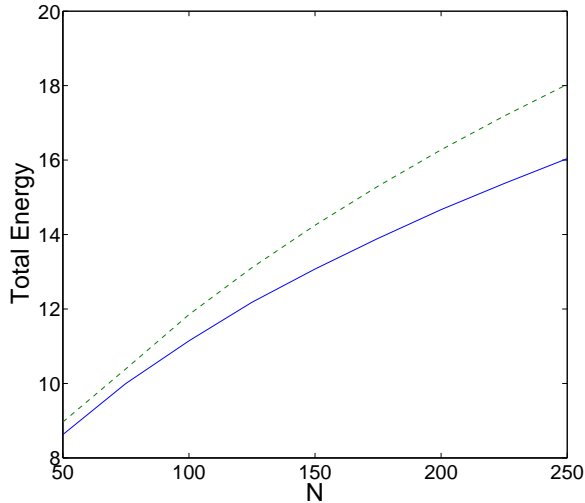,width=1.0\linewidth}}
\caption{Total energies (in $\hbar\omega/2$ units) as a function of $N$.} 
\label{figure4}
\end{figure}

We also compare the YBF model with the conventional Gross-Piteavskii (GP) equation. This can be done by taking a contact potential $g\delta(|r-r'|)$ in Eq. (\ref{Hart2}), where $g=4\pi\hbar^{2}/m\log|2\hbar^{2}/(m\mu a^{2})|$ is the coupling parameter for the two-dimensional BEC \cite{stoof,lee,rajagopal} incorporating the {\it s-wave} scattering legth $a$. We assumed a negligible exchange correlation between the atoms ($V_{xc}=0$) and would like to stress here that the condensate density is approximated to the total density $n_{GP}(r)$ at absolute zero temperature (neglecting quantum fluctuation). Based on the similar arguments in obtaining Eq. (\ref{Kohn-Sham}), we obtain the time independent Gross-Pitaevskii equation :

\begin{equation}
 \left[-\frac{\nabla^{2}}{2}+ x^{2}+ \tilde{g}n_{GP}(x) \right ]\Psi(x)=\mu\Psi(x) \, , 
\label{GP}
\end{equation}

where $\tilde{g}$ is the dimensionless scaled coupling parameter. We have also calculated the total energy of YBF and GP model as a function of the number of atoms and the result is shown in Figure (\ref{figure4}). Increasing $N$ we observe an increase of both interaction and harmonic oscillator potential energy for both GP and YBF models. The latter effect follows from the expansion of the cloud. On the contrary, the kinetic energy per particle decreases because the density profile is flattened. The GP energy curve (dash-line) remains well above YBF curve (solid line) for all ranges of $N$. The energetic superiority of GP solution demonstrate the strongly repulsive nature of delta potential compared to the soft-core nature of Yukawa potential as the number of atoms increases in the system. 
   
In summary, we have studied the system of N Yukawa bosons in a 2D harmonic trap at absolute zero temperature. The central issue of this work is the use of Density Functional Theory formalism within the Kohn-Sham scheme in reproducing the result of Monte-Carlo simulation of Magro and Ceperley \cite{magro} for the liquid phase in a harmonic trap. Physically sensible result 
through the Local Density Approximation is obtained for the trapped system by knowing the homogeneous exchange correlation energy. The ground state properties (density profiles and total energies) has been obtained by solving the non-linear equations Eqs. (\ref{Kohn-Sham}) and (\ref{GP}). Our results shows that bosons interacting with Yukawa potential is energetically favourable compared to the contact delta potential for all ranges of $N$ considered in this work. The results have so far not been verified. In light of the above, we trust that the YBF model through an appropriate mapping (boson-vortex) can be efficiently used to study strongly correlated liquid formed at the melted phase of vortices in a harmonically trapped rotating Bose gases.  

\subsection*{Acknowledgement}

I would like to thank J. F. Corney, M. K. Olsen and A.S. Bradley for useful discussions. KKR is funded by the Faculty of Engineering, Physical Sciences and Architecture (EPSA) of the University of Queensland and ARC Centre of Excellence for Quantum-Atom Optics (ACQAO), Australia.

\end{document}